\documentclass[preprint,aps,showpacs,showkeys,preprintnumbers,amsmath,amssymb]{revtex4}
\usepackage[dvips]{graphicx}
\begin{document}
\title{Complex energy approaches for calculating isobaric analogue states}

\author  {R. Id Betan}
\affiliation{
Departamento de Qu\'imica y F\'isica, FCEIA(UNR) - Instituto de F\'isica Rosario (CONICET),
Av. Pellegrini 250, 2000 Rosario, Argentina}
\author{A. T. Kruppa}
\affiliation{
Institute of Nuclear Research of the Hungarian  Academy of Sciences,
H-4001 Debrecen, P. O. Box 51, Hungary}
\author{T. Vertse}
\affiliation{
Institute of Nuclear Research of the Hungarian  Academy of Sciences,
H-4001 Debrecen, P. O. Box 51, Hungary\\
University of Debrecen, Faculty of Informatics, H-4010 Debrecen, P. O. Box 12,
Hungary}

\date{\today}

\begin{abstract}
Two methods the complex energy shell model
(CXSM) and the complex scaling (CS) approach were used for calculating
isobaric analog resonances (IAR) in the Lane model. The IAR parameters
calculated by the CXSM
and the CS methods were checked against the parameters extracted from
the direct numerical solution of the coupled channel Lane equations (CC).
The agreement with the CC results was generally better than 1 keV 
for both methods and for each partial waves concerned. 
Similarities and differences of the CXSM and the CS methods are discussed.
CXSM offers a direct way to study the configurations 
of the IAR wave function in contrast to the CS method. 
\end{abstract}



\maketitle

\section{Introduction}
\label{sec:intro}
 
The isobaric analogue resonances (IAR) have been discovered about four 
decades ago.
They are the consequences of the approximate isospin symmetry. In the absence 
of isospin symmetry braking forces the analogue state would be degenerate with
corresponding state of the parent nucleus. The Coulomb forces and isospin
dependent interactions shift up the position of the analogue state and the
analogue state may acquire a small width and may become a resonance.

The IAR has attracted interest in the latest development of the study of the
light exotic nuclei. The unusual properties of neutron reach nuclei can provide 
insights into the nuclear structure far from the valley of stability. The
extreme neutron to proton ratios may help to understand the nuclear matter at
extreme conditions. However the experimental study of the neutron rich nuclei
around the neutron drip line are difficult. Since the IAR has essentially 
the same structure as the parent state it was suggested that instead of the 
neutron reach exotic nuclei (${\rm ^{11}Li}$, ${\rm ^{14}Be}$, ${\rm ^{7}He}$,
${\rm ^{9}He}$) 
\cite{ter97,tak01,rog03,rog04} their less exotic analogue 
states should be studied (with inverse kinematics) to gain information about 
the properties of these exotic nuclei. 

Recent developments in experimental facilities
opened the possibility for identifying  large number
of exotic nuclei. To understand the structure of these nuclei
new theoretical methods have been introduced for describing the dynamics of weakly bound
or unbound nuclei from which nucleons can be emitted.
Some of the new methods is e.g. the Shell Model in the complex energy
plane (CXSM) \cite{rsm,cxsm} 
or the Gamow Shell Model \cite{mic,mic03,mic07,mic07a,rot06} 
use the Berggren basis \cite{b68}.
In the Berggren basis bound and resonant states are treated on equal footed
and scattering states taken along a contour $L$ of the complex energy sheet
are included as it will be discussed later in detail. In the last few years
this basis has been used successfully in a serious of works
\cite{bet06,dus07,ve95,blo96,li96}. Since the extended use of this
basis started not very long ago therefore we think that it is worthwhile to
accumulate more experience concerning its accuracy and its dependence on
their parameters.

Another well established method for calculating resonances is the complex
scaling (CS)
method. CS has a strict mathematical foundation given in
Refs.\cite{ABC1,ABC2,ABC3}. The possible applications and the details of the CS method
are reviewed in \cite{ho83,moi98}.

The IAR states phenomenologically can be
described by the Lane equations \cite{lan} or they can be studied by
microscopically \cite{col98}. Coupled channel (CC) Lane equations offer a 
simple but not trivial (the simplest multichannel) example on which both
 the CXSM and the CS approaches can be checked.

 Our aim in this work is to compare the parameters of the IAR calculated
by different methods. We shall compare the characteristic
features of the two methods working on the complex energy plane.
The conclusion of such a methodical work can be useful later in analyzing
experimental data in more realistic calculations. 

 The CS can be applied only for dilatation analytic 
potentials and interactions.
However some of the widely used potentials in nuclear physics are not dilatation
analytic or dilatation analytic only in a limited range of
the rotation angle, i.e. below  the critical value of the angle.
 Therefore we repeated our calculation with a
slightly modified Coulomb potential which is dilatation analytic.
This comparison is very useful to compare the accuracy of the CXSM and CS methods 
 if they give the same results
in cases when both methods can be applied.

 In section \ref{sec:lane} we summarize the features of the Lane equations.
In section \ref{sec:cxres} we describe the approximate solution of the Lane
equations using the CXSM, while in section
\ref{sec:cslane}
we make a short description of the solution of the Lane equations using
CS.
In chapter \ref{numeres} we give the numerical results of the calculations.
In the first part of chapter \ref{numeres} we compare the positions of the poles of the S-matrix 
calculated by the CXSM method with that extracted from the solution of the Lane equations.
The results of the CS method are also presented here.
The similarities and differences of the CXSM and CS methods are also discussed
in that chapter.
Finally in the last
chapter 
we summarize the main conclusions of the paper.

\section{Resonance solution of the Lane equation}
\label{sec:cxbq}
The Lane equations in the simplest case describe the quasi-elastic scattering 
of a proton and the
IAR. We assume that the target nucleus has 
mass number $A=N+Z$ and charge number $Z$. 
The ground state of the target has isospin $T_A$ and isospin projection
$T_3={\frac {N-Z}{2}}=T_A$. The target is bombarded by a beam of protons.

\subsection {Lane equation}
\label{sec:lane}
The Hamiltonian of the target plus nucleon system $H$ can be divided
to a  part describing the internal motion of the target $H(\xi)$
and their relative  motion $H_{rel}$
\begin{equation}\label{eq:hamsys}
H=H(\xi)+ H_{rel} ~.
\end{equation}
The internal state of the ground state of the target is denoted 
by $\vert A\rangle$ and this state is the solution of the equation 
$H(\xi)\vert A\rangle=\epsilon_A\vert A\rangle$. 
The analog nucleus is denoted by $\tilde A$ it has the 
same isospin $T_A$ and isospin projection $T_A-1$. It is an excited state
of the isobaric nucleus with $Z+1$ protons and $N-1$ neutrons.
If we neglect the mass difference between the neutron and proton and denote 
the additional Coulomb energy of the analogue nucleus by $\Delta_c$ 
then the eigenvalue of the internal
motion of the analogue state 
is simply $\epsilon_A+\Delta_c$ and we
have $H(\xi)\vert\tilde A\rangle=(\epsilon_A +\Delta_c)\vert \tilde A\rangle$.
Let $\vert pA\rangle$ and $\vert n\tilde A\rangle$ be the states formed by
adding a proton and neutron to $\vert A\rangle$ and $\vert \tilde A\rangle$,
respectively. The total wave function of the system may be written in the form 
\begin{equation}\label{eq:totpsi}
\Psi=\vert A\rangle \phi _p({\bf r}) +\vert \tilde A\rangle \phi _n({\bf r}),
\end{equation}
where $\phi_p({\bf r})$ and $\phi_n({\bf r})$ describe the relative motion.
The relative motion part of the total Hamiltonian can be cast into the form 
\begin{equation}\label{eq:relham}
 H_{rel}=K + V_0({\bf r}) +{\bf \hat t}\cdot{\bf \hat T} V_1({\bf r}) + ({\frac {1}{2}} -t_3) V_C(r)~,
\end{equation}
where $K$ is the kinetic energy operator of the relative motion, $V_0$ comes from the interactions
independent form the isospin, $V_C$ is the nuclear 
Coulomb potential and ${\bf \hat t}\cdot {\bf \hat T} V_1({\bf r})$ is the symmetry term due to
the isospin dependent strong interactions. 
The vector operators $\bf \hat t$ and
$\bf \hat T$ are the isospin operators of the nucleon and that of the target.
Substituting the ansatz (\ref{eq:totpsi}) into  the Schr\"odinger equation 
$H \Psi ={\cal E}\Psi$ and taking into account the form (\ref{eq:relham}) of the relative
Hamiltonian we get the Lane equations \cite{lan}:
\begin{eqnarray}
\label{laneq}
\left [ K+V_0-{{\frac {1}{2}}} T_A V_1+V_C-{\cal E}_p \right] \phi _p 
+\sqrt{{\frac{1}{2}}T_A} V_1 \phi _n =0\nonumber\\
\left[ K+V_0+{{\frac{1}{2}}} (T_A-1)V_1-({\cal E}_p-\Delta_c) \right] \phi _n 
+\sqrt{{\frac{1}{2}}T_A} V_1\phi_p=0,
\end{eqnarray}
where ${\cal E}_p={\cal E}-\epsilon_A$ is the center-of-mass energy of the relative motion of the 
proton plus target system and the relative energy for the neutron plus analogue nucleus is 
${\cal E}_p-\Delta_c$. If we assume that the interactions are spherical
symmetric then 
the relative motion
can be divided into partial waves and there are no coupling between different
partial waves characterized by orbital $l$ and total angular momentum $j$
quantum numbers.
The numerical solution of the Lane equation  has 
been carried out by using fourth order Runge-Kutta method.
At each real ${\cal E}_p$ values we
calculated two linearly independent solutions of the coupled equations.
The physical solution with components $\phi _p$  and $\phi _n$ being regular at $r=0$ was combined from these
independent solutions. These components $\phi _p$  and $\phi _n$
were matched to the scattering (or outgoing wave) solutions of the corresponding channels
at a distance where the nuclear potentials are cut to zero.
This is a standard method described e.g. in Ref. \cite{toboc}.

\subsection{CXSM a solution using Berggren basis}
\label{sec:cxres}
In this section we calculate the complex energy eigenvalues of the IAR by
diagonalizing the Hamiltonian (\ref{eq:hamsys}) in combined Berggren 
bases \cite{b68} of the target plus proton and
analog plus neutron systems. First we describe the Berggren basis for the protons.
We consider an auxiliary problem, a radial Schr\"odinger equation with the diagonal potential
of the first equation of the  Lane equation (\ref{laneq}) 
\begin{equation}
\left [ K_l+V_0-{{\frac{1}{2}}} T_A V_1+V_C-E_n^{(p)} \right] u^{(p)}_n(r)
=0,
\end{equation}
where
\begin{equation}
K_l=-{\frac{\hbar^2}{2\mu}}  \left [{\frac{d^2}{dr^2}} -{\frac{l(l+1)}{r^2}} \right] 
.
\end{equation}

The discrete bound and resonance solution with energy $E_n^{(p)}$ are 
denoted by $u^{(p)}_n(r)$ and the scattering solutions by 
$u^{(p)}(r,E)$. Sometimes when it is obvious we will use the 
wave number $k$ instead of the energy $E$ and the scattering states 
along the contour in the
lower half of the second energy sheet will be denoted by 
$u^{(p)}(r,k)$.
 
\begin{figure}
\begin{center}
\begin{picture}(250,200)(-120,-100)
\thinlines
\put(-120,0){\vector(1,0){250}}
\put(0,-100){\vector(0,1){200}}
\thicklines
\put(0,0){\line(2,-1){80}}
\put(80,-40){\line(1,1){40}}
\put(0,0){\line(-2,1){80}}
\put(-80,40){\line(-1,-1){40}}
\put(0,18){\circle{3}}
\put(0,50){\circle{3}}
\put(40,-10){\circle{3}}
\put(62,-20){\circle{3}}
\put(4,98){\makebox(0,0)[lt]{$\Im (k)$}}
\put(132,4){\makebox(0,0)[rb]{$\Re (k)$}}
\put(-4,18){\makebox(0,0)[r]{$b_2$}}
\put(-4,50){\makebox(0,0)[r]{$b_1$}}
\put(42,-12){\makebox(0,0)[lt]{$d_1$}}
\put(66,-23){\makebox(0,0)[lt]{$d_2$}}
\put(80,-45){\makebox(0,0)[l]{$L^+$}}
\put(-80,50){\makebox(0,0)[l]{$L^-$}}
\end{picture}
\end{center}
\caption {\label{path}
{Positions of the bound ($b_1$, $b_2$) and decaying resonant ($d_1$, $d_2$) poles of $S(k)$ on the complex
$k$-plane and a possible choice of the complex contour
$L$~.}}
\end{figure}
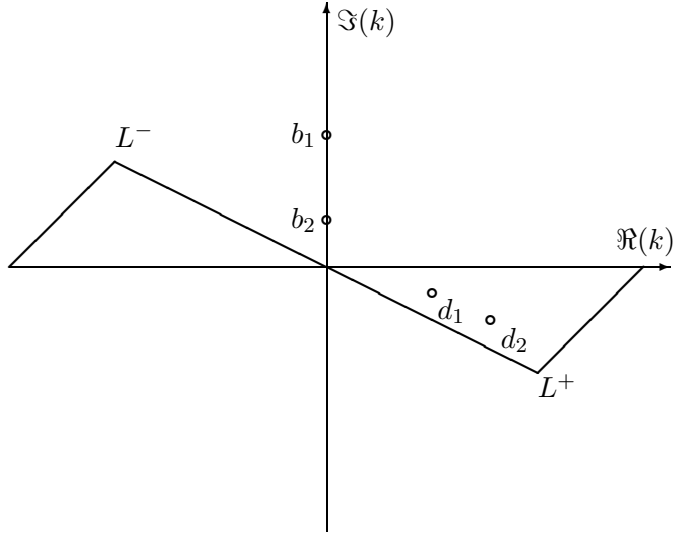

The main advantage of the Berggren
basis is that the single-particle basis set consists not only of bound states 
but also of poles of the single-particle Green
function on the complex energy (wave number) planes and a continuum
of scattering states taken along a complex contour $L$. 
A typical contour of the complex wave number plane
is shown in Fig. \ref{path}.
The $L^+$ part of the contour goes from the origin to infinity in the
lower half of the second energy sheet, while the $L^-$ part of the contour
makes exactly the same tour on the first energy sheet. 
It was observed in Refs. \cite{Hag06,Myo98} that the contour  $L^+$ need not return to the
real axis at infinity.
The shape of the chosen complex contour $L=L^+
+ L^-$ regulates
which of the poles should be included into the Berggren basis forming the 
 completeness relation of Berggren: 

\begin{equation}\label{eq:delb}
 \delta(r-r^\prime)=\sum_{n=b,d}{u^{(p)}_n(r)}u^{(p)}_n(r^\prime)
 +2\int_{L^+} dk {u^{(p)}(r,k)}u^{(p)}(r^\prime,k)~.
\end{equation}
In this relation (and later) the notation $n=b,d$ means that the sum over 
$n$ runs through all  bound states plus 
the decaying resonances 
lying between the real energy axis and the
integration contour $L^+$ of Fig. \ref{path}. The integral in 
Eq. (\ref{eq:delb}) is over the
scattering states along $L^+$ (factor $2$ is due to the symmetry of $L$). 
The poles denoted by $d$ in the basis generally correspond to  decaying
resonances lying
on the fourth quadrant of the complex $k$-plane. 

The completeness relation in Eq. (\ref{eq:delb}) was introduced for charge less particles in Ref. \cite{b68}
and it has been shown later in Ref. \cite{Michel} that it is valid even for charged particles.
Berggren completeness can be generalized by using a contour of different shape
in which  
antibound states \cite{i05} lying on the negative part of the imaginary 
$k$-axis are included in the sum in 
Eq. (\ref{eq:delb}). Since the inclusion of antibound states
is not optimal as far as the number of basis states is concerned \cite{Naza}
we are not using antibound basis states in this work. 

Berggren introduced a generalized scalar product between functions defining 
a special  complex metric of
the Berggren space \cite{b68}.
In the generalized scalar product in the left (bra) position of the scalar product the
mirror partner state (denoted by tilde over the state) is used. This state corresponds to the reflection to the
imaginary $k$-axis. Due to this reflection
in this scalar product in the integration 
the radial wave functions appears instead of the complex conjugate of the
radial function. (This causes no difference for bound (and anti-bound) states
lying on the imaginary axis.) 
This is the only modification in the scalar product since the spin-angular degrees of freedom
remains unchanged. If the radial integral to be calculated has no definite value then a regularization
procedure has to be applied. Zel'dovich \cite{zel} and also Romo \cite{rom} suggested regularization
methods but we use the complex rotation of the radial distance $r$ beyond the
range of nuclear forces \cite{gyar}.

The upper half of the complex $k$-plane maps to the physical (or first)
Riemann-sheet of the complex energy $E\sim k^2$. 
The pole wave functions of this sheet
are square integrable functions belonging to bound states.
While the lower half  of the complex $k$-plane maps to the unphysical (or second)
Riemann-sheet of the energy $E$. The pole wave functions of the second sheet
are not square integrable functions and they belong to decaying/capturing
resonances lying on the lower/upper part of that energy sheet or
antibound states lying on the negative real energy axis. 
The calculation of integrals in which these
radial wave functions appear might need the use of regularization procedure. 

Since the number of basis states has to be finite  the complex continuum has to be 
discretized. It is preferable to use as discretization points $E_i^{(p)}$ the abscissas of a Gaussian
quadrature procedure and the corresponding weights of that procedure
are denoted by $h_i$. By discretizing the integral in Eq. (\ref{eq:delb}) one 
obtains an approximate completeness relation for the finite number of basis
states:
\begin{equation}\label{eq:finb}
 \delta(r-r^\prime)\approx \sum_{n=b,d,c}^M {w^{(p)}_n(r,E^{(p)}_n)}w^{(p)}_n(r^\prime,E^{(p)}_n),
\end{equation}
where $c$ labels the discretized contour $L^+$ states. 
If $E^{(p)}_n$ corresponds to scattering energy from the contour $L^+$ then 
the scattering state of the discretized continuum is denoted by 
$w^{(p)}_n(r,E^{(p)}_n)=\sqrt{h_n} u^{p}_n(r,E^{(p)}_n)$ and if $E^{(p)}_n$ 
corresponds to a normalized pole state then 
$w^{(p)}_n(r,E^{(p)}_n)= u^{(p)}_n(r)$. 
The set of Berggren vectors form a bi-orthonormal basis in the truncated space
\begin{equation}\label{eq:biort}
 <\tilde {w}^{(p)}_n|w^{(p)}_m>=\delta_{n,m}~.
\end{equation}
The Berggren basis for  neutrons is defined similarly but the auxiliary
problem uses the diagonal part of the second equation of (\ref{laneq})
\begin{equation}
 \left[ K_l+V_0+{{\frac{1}{2}}} (T_A-1)V_1-E^{(n)}_n \right] u^{(n)}_n({r}) =0.
\end{equation}

Having fixed the Berggren basis for neutrons and protons 
we take the ansatz (\ref{eq:totpsi}) but 
the relative motion functions are expanded on the corresponding 
Berggren basis

\begin{equation}
\phi_p({\bf r})=\left [ \sum_{i=1}^{M_p}C^{(p)}_iw^{(p)}_i({
r},E^{(p)}_i)\right ] {\cal Y}_{ljm}
\end{equation}
and
\begin{equation}
\phi_n({\bf r})=\left [\sum_{i=1}^{M_n}C^{(n)}_iw^{(n)}_i({r},E^{(n)}_i)\right
]{\cal Y}_{ljm}~,
\end{equation}
where ${\cal Y}_{ljm}$ denotes the spin-angular part of the wave function.
Using Eq. (\ref{laneq}) we get the following set of linear equations for the 
unknown complex expansion coefficients 
$C^{(p)}_i$ and $C^{(n)}_i$
\begin{eqnarray}\label{eq:secul1}
&(E^{(p)}_k-{\cal E}_p)C^{(p)}_k+\sum_{m=1}^{M_n}\langle\tilde w^{(p)}_k|\delta
v|w^{(n)}_m\rangle C^{(n)}_m=0\nonumber \\&k=1,\ldots,M_p
\end{eqnarray}
and
\begin{eqnarray}\label{eq:secul2}
&(E^{(n)}_k-({\cal E}_p-\Delta_c))C^{(n)}_k+\sum_{m=1}^{M_p}\langle\tilde w^{(n)}_k|\delta
v|w^{(p)}_m\rangle C^{(p)}_m=0\nonumber\\ &k=1,\ldots,M_n,
\end{eqnarray}
where the coupling potential is $\delta v= \sqrt{\frac{T_A}{2}} V_1$.
The above two equations can be combined into one matrix eigenvalue equation with dimension 
$M_p+M_n$. By diagonalizing the matrix of the Hamiltonian  we get $M_p+M_n$ complex 
eigenvalues ${\cal E}^\nu_p$ $\nu=1,\ldots M_p+M_n$.
One of the complex eigenvalues ${\cal E}^\nu_p$ is identified by the energy of the IAR.
The identification  in general is easy because most of the other unbound states
correspond to discretized contour states and they are lying far from the
position of the IAR at 
${\cal E}_{IAR}=E_r-i{\frac{\Gamma}{2}}$ and  in the wave function
 of the IAR the dominant component is  a bound neutron state. 

\subsection{Complex scaled Lane equation}
\label{sec:cslane}

The poles of the Green-operator on the complex energy plane can be determined with the help of
the complex scaling. The CS mathematically well founded
\cite{ABC1,ABC2,ABC3} and has many applications in atomic, molecular and nuclear
physics. 
We demonstrate the effect of the CS  on an example of a
single-particle Hamiltonian $\hat h$. The real angle $\theta$ of the 
CS rotates the coordinates of the particle to complex, i.e.
$\bf r$ is simply replaced by $\exp(i\theta){\bf r}$. More precisely
the effect of the CS can be given with the help of an operator 
${\hat U}(\theta)$. 
It acts on an  arbitrary function $g({\bf r})$ as 
\begin{equation}
  {\hat U}(\theta)g({\bf r})=\exp(i{\frac{3} {2}}\theta)g({\bf r}e^{i\theta})~.
\end{equation}
The complex scaled Hamiltonian is the following
\begin{equation}
 {\hat h}_\theta={\hat U}_\theta{\hat h}{\hat U}_\theta^{-1}~.
\end{equation}
The kinetic energy $\hat{K}=-\frac{\hbar^2}{2 \mu} \Delta_{\bf r}$ transforms due to the complex scaling to

\begin{equation}
 {\hat U}_\theta{\hat K}{\hat U}_\theta^{-1}=\exp(-i2\theta)(-\frac{\hbar^2}{2 \mu} \Delta_{\bf r})~,
\end{equation}
and a local potential ${\hat V}(\bf r)$ transforms to the form:
\begin{equation}
 \hat V^\theta({\bf r})={\hat U}_\theta{\hat V}({\bf r}){\hat U}_\theta^{-1}={\hat V}({\bf
 r}\exp(i\theta))~.
\end{equation}
Assume that $\chi_\nu(\bf r)$ is a 
bound or resonance eigenfunction of the the Hamiltonian $\hat h$ and the corresponding 
eigenvalue is $E_\nu $ then the function 
$\chi_\nu^\theta ({\bf r})=\exp(i{\frac{3}{2}}\theta)\chi_\nu ({\bf r}e^{i\theta})$ will be the eigenfunction of 
the complex scaled Hamiltonian ${\hat h}_\theta$ with the same eigenvalue $E_\nu $.
The advantage of the CS is that the function $\chi_\nu^\theta ({\bf r})$ is square integrable even if
the original state was a resonance wave function. The square integrability of 
$\chi_\nu^\theta ({\bf r})$ allows that it can be approximated well with finite expansion using only
square integrable basis functions.

The Lane equation can be considered as an eigenvalue problem of a two by two matrix Hamiltonian

\begin{equation}
 {\cal H}=
 \left (
 \begin{array}{cc}
  K+V_0-{\frac{T_A}{2}}  V_1+V_C&\ \ \sqrt{\frac{T_A}{2}} V_1\\ 
  \sqrt{\frac{T_A}{2}} V_1 &\ \   K+V_0+{\frac{T_A-1}{2}} V_1+\Delta_c
 \end{array}\right)~.
\end{equation}
The Lane-equation (\ref{laneq}) can be cast into the form 
\begin{equation}
{\cal H}
\left (\begin{array}{c}
\phi_p({\bf r})\\
\phi_n({\bf r})
\end{array}\right)={\cal E}_p\left (\begin{array}{c}
\phi_p({\bf r})\\
\phi_n({\bf r})
\end{array}\right).
\end{equation}
The generalization of the operator $\hat U_\theta$ is straightforward 
\begin{equation}
{\cal U}_\theta=\left( \begin{array}{cc}
U_\theta &
0\\
0&U_\theta
\end{array}\right)
\end{equation}
and the complex scaled matrix Hamiltonian is 
${\cal H}_\theta={\hat {\cal U}}_\theta
{\cal H}{\hat {\cal U}}_\theta^{-1}$. 
The eigenvalue problem of this operator
\begin{equation}
 {\cal H}_\theta \left (\begin{array}{c}
 \phi^\theta_p({\bf r})\\
 \phi^\theta_n({\bf r})
 \end{array}\right)={\cal E}^\theta_p\left (\begin{array}{c}
 \phi^\theta_p({\bf r})\\
 \phi^\theta_n({\bf r})
 \end{array}\right)
\end{equation}
in components gives the following set of equations
\begin{eqnarray}
 \label{laneqcs}
 \left [H_p^\theta-{\cal E}^\theta_p \right]
  \phi^\theta _p 
 +\sqrt{\frac{1}{2}T_A} V_1^\theta \phi^\theta_n =0\nonumber\\
 \left[ H^\theta_n-{\cal E}^\theta_p \right]
  \phi^\theta_n
 +\sqrt{\frac{1}{2}T_A} V_1^\theta\phi^\theta_p=0 ~,
\end{eqnarray}
where $H^\theta_p=\exp(-i2\theta)K+V^\theta_0-\frac{1}{2} T_A V^\theta_1+ V^\theta_C$ 
and $H^\theta_n= \exp(-i2\theta) K+V^\theta_0+\frac{1}{2} (T_A-1)V^\theta_1+\Delta_c$. 
We will refer to (\ref{laneqcs}) as complex scaled Lane-equation. 
Since the functions $\phi^\theta_p({\bf r})$ and $\phi^\theta_n({\bf r})$ are square integrable we can make the approximation
 
\begin{equation}
 \phi^\theta_p({\bf r})=\left [\sum_{i=1}^{M_p}C^{(p,\theta)}_i\psi_i^{(p)}({r})\right ] {\cal Y}_{ljm}
\end{equation}
and
\begin{equation}
 \phi^\theta_n({\bf r})=\left [\sum_{i=1}^{M_n}C^{(n,\theta)}_i\psi_i^{(n)}({r})\right ] {\cal Y}_{ljm},
\end{equation}
where $\psi_i^{(n)}({r})$ and $\psi_i^{(p)}({r})$ are arbitrary square integrable basis functions.
Substituting these forms into the (\ref{laneqcs}) we get a matrix eigenvalue equation. In
details we have
\begin{eqnarray}\label{eq:cs1}
 \sum_{m=1}^{M_p}\langle\tilde \psi_k^{(p)}|
 H^\theta_p|\psi_m^{(p)}\rangle C^{(p,\theta)}_m+
 \sum_{m=1}^{M_n}\langle\tilde \psi_k^{(p)}|
 \delta v^\theta|\psi_m^{(n)}\rangle C^{(n,\theta)}_m\nonumber\\
 ={\cal E}_p\sum_{m=1}^{M_p}\langle\tilde \psi_k^{(p)}|\psi_m^{(p)}\rangle C^{(p,\theta)}_m\ \ \ k=1,\ldots,M_p,
\end{eqnarray}
and
\begin{eqnarray}\label{eq:cs2}
 \sum_{m=1}^{M_n}\langle\tilde \psi_k^{(n)}|
 H^\theta_n|\psi_m^{(n)}\rangle C^{(n,\theta)}_m+
 \sum_{m=1}^{M_p}\langle\tilde \psi_k^{(n)}|
 \delta v^\theta|\psi_m^{(p)}\rangle C^{(p,\theta)}_m\nonumber\\
 ={\cal E}_p\sum_{m=1}^{M_n}\langle\tilde \psi_k^{(n)}|\psi_m^{(n)}\rangle C^{(n,\theta)}_m
 \ \ \ k=1,\ldots,M_n.
\end{eqnarray}
The solution of these equations provide us $M_p+M_n$ number of complex
eigenvalues. The majority of these eigenvalues correspond to the 
discretization of the rotated down continuous spectrum. The bound and resonance
poles can be clearly identified and the accurate value can be determined using 
the so called $\theta$ trajectory technique.

\section {Numerical results}
\label{numeres}
We applied the methods described in section \ref{sec:cxres} and \ref{sec:cslane} for the description of the IAR-es
in the ${\rm ^{209}Bi}$  nucleus with large neutron excess. We studied several analogue resonances in the ${\rm p+^{208}Pb}$ system.
For illustrative purposes we selected an IAR which in our simple model is the analog of the ground state of ${\rm ^{209}Pb}$, i.e.
a $g_{9/2}$ single particle state.
The effect of the double magic core is described by a phenomenological potential. We used Woods-Saxon (WS) forms for
both the diagonal and the coupling potentials in (\ref{laneq}).
The WS forms cut
to zero at a finite distance: $R_{max}=20$ fm
\begin{equation}
\label{vagottWS}
V^{WS}_{tr}(r)=\left\{
\begin{array}{rl}
V^{WS}(r) &\textrm{ if } r~<~R_{max}\\
0&\textrm{ if } r~\geq~ R_{max}~.
\end{array}
\right.
\end{equation}
The spin-orbit part of the potential has the usual derivative form:
\begin{equation} \label{vso}
 V_{so}^{WS}(r)=-\frac{V_{so}}{r a}~ 2(\vec l \cdot \vec s)
   \frac{e^{\frac{r-R}{a}}}{(1+e^{\frac{r-R}{a}})^2}~.
\end{equation}
It is also cut to zero at $R_{max}$.
The numerical values of the potential parameters were taken from an early
work \cite{gyave}.
For the sake of simplicity the radii and the diffuseness were taken the same
values for protons and neutrons and for the common spin-orbit term:
$r_0=1.19$ fm and $a=0.75$ fm, $V_{so}=11.6$ MeV. 
For Coulomb potential we assumed that the charge $Ze$ of the target is
homogeneously distributed inside a sphere with radius
 $R_c=r_c A^{1/3}$ with sharp edge
\begin{equation}
 \label{vcou}
 V_C(r)=Ze^2 \left\{
 \begin{array}{rl}
  \frac{1}{2R_c}[3-(\frac{r}{R_c})^2]
 & \textrm{ if  } r \le R_c\\ 
 1/r&\textrm{ if } r>R_c ~.
 \end{array}
 \right.
\end{equation}
The depth of the nucleon potential was $56.4$ MeV and the strength of the
symmetry potential was $0.5$ MeV.
Therefore the diagonal WS potential felt by the proton was $61.9$ MeV and by the
neutron $51.15$ MeV
according to the Lane equations in Eq. (\ref{laneq}). The Coulomb radius was
identical with the one of the nuclear potential.
The Coulomb energy difference was also the same as in Ref. \cite{gyave} $\Delta_c=18.9$ MeV.

\begin{table}
\begin{ruledtabular}
\begin{tabular}{|c|c|c|c|}
State&neutron&proton(Eq. (\ref{vcou}))&proton(Eq. (\ref{dilcoul}))\\
\hline
$1g_{9/2}$&$(-22.878,0.0000)$&$(-11.894,0.0000)$&$(-13.975,0.0000)$\\
$2g_{9/2}$&$(-4.060,0.0000)$&$(7.674,-6.2 10^{-4})$&$(6.070,-2.0 10^{-5})$\\
\end{tabular}
\end{ruledtabular}
\caption{Energies of the discrete $g_{9/2}$ single-particle basis states for neutrons
and for protons 
 corresponding to the Woods-Saxon
potential described in the text. The Coulomb potential for protons
is either the usual one in Eq. (\ref{vcou}) or the dilatation analytic one in
Eq. (\ref{dilcoul}).
 Energies are in MeV. \label{spener}}
\end{table}

In the CXSM the elements of the single particle bases
are calculated in the diagonal potentials appearing in the corresponding
channels of the Lane equations.
The single particle energies for the $g_{9/2}$ neutron and proton orbits are summarized
in Table \ref{spener}.
The vertexes of three different proton and neutron $L$ contours for the $g_{9/2}$ case are shown in Table
\ref{cont}.
The numbers of the discretization points $N_i$ of the segment $[V_i,V_{i+1}]$  are shown between the vertex points.
To calculate IAR-es we could use neutron contours taken along the real axis.
For other partial waves we used large variety of contours.

\begin{table}[ht]
\begin{ruledtabular}
\begin{tabular}{|c|c|c|c|c|c|c|}
Channel&\multicolumn{3}{c|}{proton}&\multicolumn{3}{c|}{neutron}\\
\hline
Contour&$LP1$&$LP2$&$LP3$&$LN1$&$LN2$&$LN3$\\
\hline
$V_0$&(0,0)&(0,0)&(0,0)&(5,-0.4)&(5,-0.35)&(3,0)\\
$N_0$&0&0&4&0&0&10\\
$V_1$&(5,-0.4)&(5,-0.35)&(5,-0.4)&(30,-0.4)&(30,-2.098)&(3,-10)\\
$N_1$&78&34&22&0&0&4\\
$V_2$&(30,-0.4)&(30,-2.098)&(30,0)&(30,0)&(100,-6.993)&(10,-10)\\
$N_2$&2&4&2&0&0&6\\
$V_3$&(30,0)&(100,-6.993)&(100,0)&(100,0)&&(10,0)\\
$N_3$&0&&4&0&0&4\\
$V_4$&(100,0)&&(200,0)&(200,0)&&(30,0)\\
$N_4$&0&&&0&0&0\\
$V_5$&(200,0)&&&&&(100,0)\\
	\end{tabular}
\end{ruledtabular}
\caption{Integration contours for $g_{9/2}$ protons and neutrons given by
vertexes $V_i$ (in MeV) and the number of Gaussian points
$N_i$. The $N_i$ values are the ones necessary to reach
the $1$ keV accuracy for the IAR or for the
broad resonance at ${\cal E}_p=(23.996,-6.147)$ MeV. Contours LP3 and LN3 were used for the broad resonance.
\label{cont}} 
\end{table}

In the CC method we solved the coupled Lane equations for a fine equidistant mesh of the
bombarding
proton energy  ${\cal E}_p={\cal E}_0,{\cal
E}_0+d{\cal E},...,{\cal E}_{max}$ in the center of mass system and calculated the
scattering matrix elements for each energy values: $S({\cal E}_p)$. 

In order to determine the parameters of the IAR we fitted the tabulated values
of the
$S({\cal E}_i)$ by the following form:
\begin{equation}
\label{poleform}
S({\cal E}_p)=e^{2i\delta_p({\cal E}_p)}(1-i\frac{\Gamma_p}{{\cal E}_p-
\cal{E}_{IAR}})~,
\end{equation}
where $\cal{E}_{IAR}$ denotes the complex energy of the
IAR, i.e. ${\cal E}_{IAR}=E_r-i{\frac{\Gamma}{2}}$, $\Gamma$ is the full width and
$\Gamma_{p}$ is the proton partial width of the IAR. Below the threshold of the
$^{208}Pb(p,{\bar n})^{208}Bi$ reaction the total width is equal to the
the partial width: $\Gamma=\Gamma_{p}$  in our model. Equation (\ref{poleform}) represents a one
pole approximation to the S-matrix in the proton channel.

For the background phase shift in the entrance channel $\delta_p({\cal E}_p)$ 
we take a linear energy dependence in order
to better reproduce the non-resonant background. Naturally all these quantities
refer to definite ${\rm l,j}$ partial waves.
The best fit parameter values  are listed in the $E_r(CC)$ and $\Gamma(CC)$
columns in Table \ref{compare}. The one pole formula of (\ref{poleform}) 
gave excellent fit to the tabulated values of $S({\cal E}_i)$ in all cases in Table  \ref{compare}.

\begin{table}[ht]
\begin{ruledtabular}
\begin{tabular}{|c|cc|cc|}
$l~j$&$E_r (CXSM)$&$E_r (CC)$&$\Gamma (CXSM)$&$\Gamma (CC)$\\
\hline
$g_{9/2}$&14.954&14.954&0.046&0.047\\
$i_{11/2}$&15.526&15.526&0.003&0.003\\
$d_{5/2}$&16.445&16.444&0.141&0.140\\
$s_{1/2}$&16.918&16.917&0.156&0.156\\
$g_{7/2}$&17.367&17.367&0.086&0.084\\
$d_{3/2}$&17.441&17.440&0.144&0.145\\
$j_{15/2}$&18.774&18.774&0.006&0.006\\
\end{tabular}
\end{ruledtabular}
\caption{Comparison of the IAR parameters calculated by using the CC and CXSM 
methods for different partial waves with Coulomb potential Eq. (\ref{vcou}). Energies are in MeV units.
\label{compare}}
\end{table}

The numerical values of ${\cal E}_{IAR}$ are shown as $E_r(CXSM)$ and 
$\Gamma(CXSM)$
in Table \ref{compare} in comparison with the results extracted from the 
solution
of the coupled Lane equations $E_r(CC)$ and 
$\Gamma(CC)$.
As one can see from the comparison the positions and the widths calculated by
the CXSM agree well (within $1$ keV) with the result of the CC.

Let us discuss below briefly how this agreement has been achieved. 
We optimized the shape of the contours and the number of
points along the contours separately for neutrons and protons and for different
partial waves. 
The shape of the contour is fixed by the vertexes which were chosen to
be able to include the narrowest single-particle resonant states.
 We observed that the contour should not go close neither to the
energies of the resonances included into the basis nor the IAR resulted by the
diagonalization.
The last vertex point i.e. the energy of the last segment with $N_i\neq 0$ was crucial to get
good agreement for both the real and the imaginary parts of the IAR energy
calculated by solving the Lane-equation.

We tested the convergence of the IAR energies by increasing the number of discretization
points and stopped to increase it when the energy did not change. After that
 we continued with the
next interval and increased the points of that interval similarly.
After going through all the intervals we optimized the number of mesh-points
by reducing them until the energy in keV did not change. 
We also tested the convergence of the IAR energies by varying the positions of the
vertexes.
If the contour goes very far from the real axis i.e. if we choose the 
value of the imaginary parts of the vertexes considerably larger than the ones
in Table \ref{cont} then the degree of agreement might be spoiled even if we
choose larger number of discretization points.
We found for all partial waves that the IAR resonances are not very much 
sensitive to the low energy part of the continuum (below $5$ MeV), neither for neutron nor for 
proton. At high 
energy however a cutoff smaller than $30$ MeV affects the convergence of the 
IAR energy.

In order to be able to compare pole solutions
for calculation of the resonance parameters of the IAR we repeated the
calculation by applying the CS for the solution of the
Lane equations. Unfortunately the Coulomb potential of a charged sphere
with sharp edge is not dilatation analytic because this form
becomes discontinuous for $\theta\ne 0$. We used the Coulomb potential
expressed by the error function which is dilatation analytic. This form 
of the Coulomb potential
\begin{equation}
\label{dilcoul}
 V_C(r)= Ze^2~\frac{{\rm Erf}(r/\alpha)}{r}
\end{equation}
is widely used in 
both atomic and nuclear physics \cite{Myo98,tou05}. In the resonating group 
model it can be obtained as the direct folding interaction 
between nuclei \cite{sai77}. The numerical value of the parameter $\alpha=0.31$ fm was adjusted to the
Coulomb potential in Eq. (\ref{vcou}).
For the nuclear potential we kept the WS form which is dilatation analytic
until the rotation angle is below  the critical angle:
$\theta<\theta_{crit}={\rm arctg} ({\frac{a\pi}{R}})$.

For the solution of the complex scaled Lane equation we used the Laguerre mesh 
basis functions 
\begin{equation}
\psi_i^{(\nu)}({r})=(-1)^ir_i^{-1/2}\frac{r L_{M_\nu}(r)}{r-r_i}\exp(-r/2),
\end{equation}
where $\nu=p,n$.
The mesh points given by $L_{M_\nu}(r_i)=0$, where $L_{M_\nu}(r)$ is the 
Laguerre polynomial. The advantage of this basis is that the matrix elements 
of any local potential is extremely simple \cite{bay02}. 
This type of basis functions are proved to be very accurate both in 
simple model calculations and in three body problems \cite{bay86,bay97,bay02}. 
One can introduce an additional simple scaling parameter of the basis 
\cite{bay02} for this parameter we used the $0.3$ fm value.

The agreement between the pole positions calculated by the CXSM and the CS
method is extremely good
for all partial waves in Table \ref{comparecs}. One can see in this Table that
the agreement with the numerically exact solution of the Lane-equation (CC)
is as good as in the previous case when the CXSM method was used with the standard
Coulomb potential. The maximal difference does not exceeds $1$ keV.

\vskip 1.cm
\begin{figure}[ht]
\includegraphics[width=8.6cm]{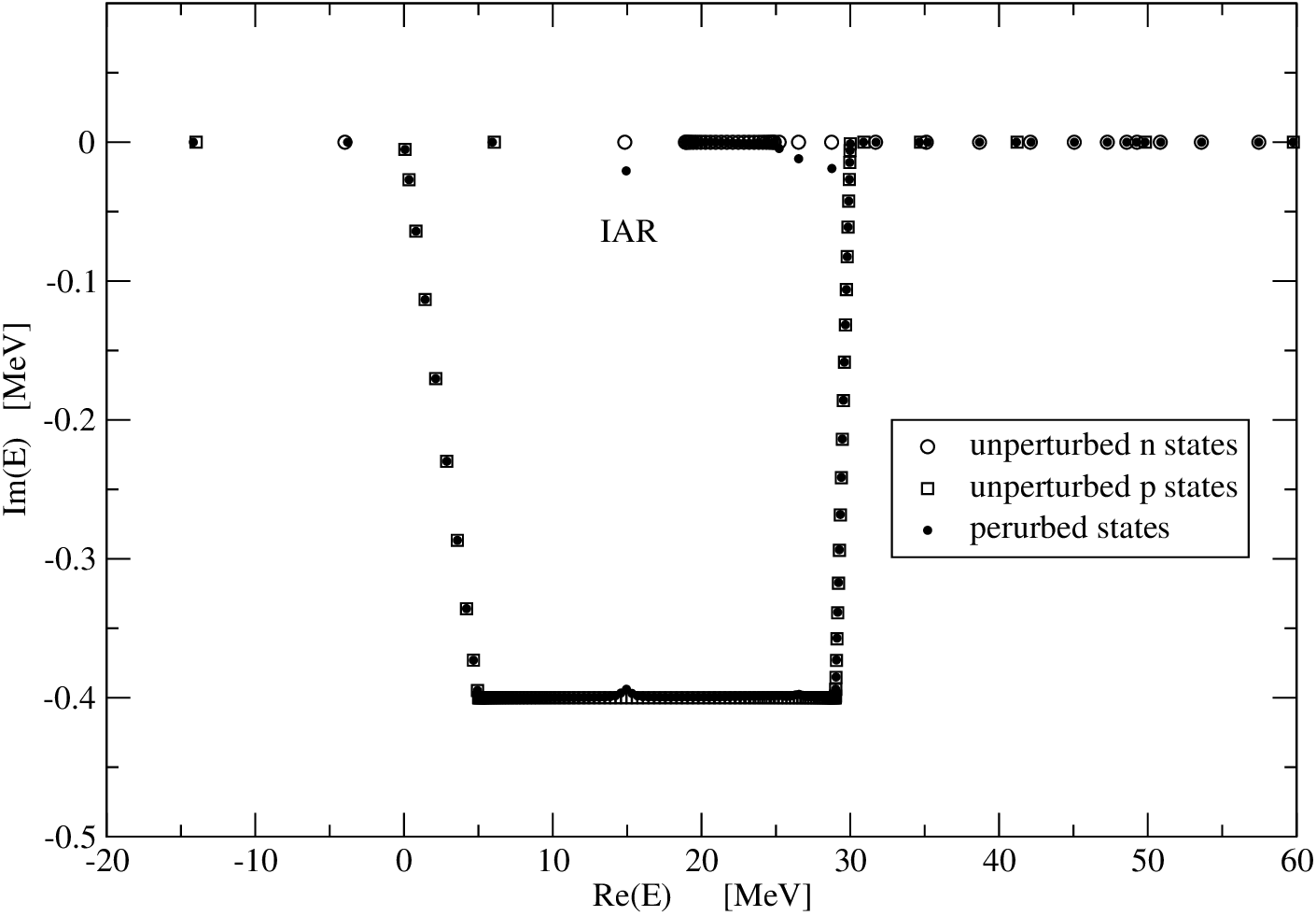}
\caption{Positions of the  $g_{9/2}$ Berggren basis states (circles for neutrons
and squares for protons) and the
results of the CXSM method (filled circles) on the
complex $E$-plane for the Coulomb potential Eq. (\ref{vcou}). }
\label{poles}
\end{figure}

To understand better the formation of the IAR let us consider again the $l=4$, $j=9/2$ case as an example. In Fig. \ref{poles}
we show the positions of the unperturbed states forming the Berggren basis
(denoted by
 circles for neutrons and squares for protons) and some of the results of the 
 diagonalization (perturbed states
 denoted by filled circles) on the
complex energy plane. A section of the real $E$-axis and of the lower half
of the complex plane is shown. In this case the neutron contour is along the
real axis while the proton contour has a trapezoidal shape with vertexes denoted by LP1
in Table \ref{cont}. In order to see better the region of our interest the
states with energies higher than 
 $60$ MeV are not shown in Fig.\ref{poles}. The discrete basis states are listed in Table
 \ref{spener}. The bound basis states 
are the $1g_{9/2}$ proton state at $E_i^{(p)}=-13.975$ MeV and the two neutron
states:  $1g_{9/2}$, $2g_{9/2}$
at $E_i^{(n)}=-22.878$ MeV and at $E_i^{(n)}=-4.060$ MeV which are shifted up by
$\Delta_c=18.9$ MeV . The narrow $2g_{9/2}$ proton resonance at
$E_i^{(p)}=(6.070,-2\times 10^{-5})$ MeV seems to lie on the real axis.

\begin{table}
\begin{ruledtabular}
\begin{tabular}{|c|cc|cc|cc|}
$l~j$&$E_r (CXSM)$&$E_r (CS)$&$E_r (CC)$&$\Gamma (CXSM)$&$\Gamma (CS)$&$\Gamma (CC)$\\
\hline
$g_{9/2}$&14.933&14.933&14.934&0.041&0.041&0.042\\
$i_{11/2}$&15.493&15.493&15.493&0.002&0.002&0.002\\
$d_{5/2}$&16.436&16.436&16.444&0.121&0.120&0.120\\
$s_{1/2}$&16.913&16.913&16.913&0.127&0.127&0.128\\
$g_{7/2}$&17.350&17.349&17.349&0.076&0.075&0.074\\
$d_{3/2}$&17.434&17.434&17.433&0.118&0.119&0.120\\
$j_{15/2}$&18.752&18.751&18.752&0.005&0.005&0.005\\
\end{tabular}
\end{ruledtabular}
\caption{Comparison of the IAR parameters calculated by using the CC, CXSM and CS 
methods for different partial waves with the dilatation analytic 
 Coulomb potential Eq. (\ref{dilcoul}). Energies are in MeV units.\label{comparecs}}
\end{table}

Most of the perturbed states lie close to the positions as the corresponding
basis states since the coupling symmetry potential term cases only a small shift
for these states. One of the exception is the IAR at ${\cal
E}_{IAR}=(14.933,-0.021)$
MeV which shifted down well below the bound $2g_{9/2}$ neutron state which is the
main component of its wave function $C^{(n)}_i=(0.9921,-0.0047)$.
The second largest component is that of the $2g_{9/2}$ proton resonance with
$C^{(p)}_i=(-0.1194,0.0002)$.
The other perturbed states which do not fit to the path of the contours are
states based on contour states but fall off the contour because of the finite
number of discretization points. If the number of discretization points are
increased they move closer to the contour. They move also with the contour if
we change the shape of the contour in contrast to the IAR which remains in the
same position. Of course the IAR should lie above the proton contour in order
to be explored. This feature is very similar to the one observed in the CS 
calculation. 

\begin{figure}[ht]
\includegraphics[width=8.6cm]{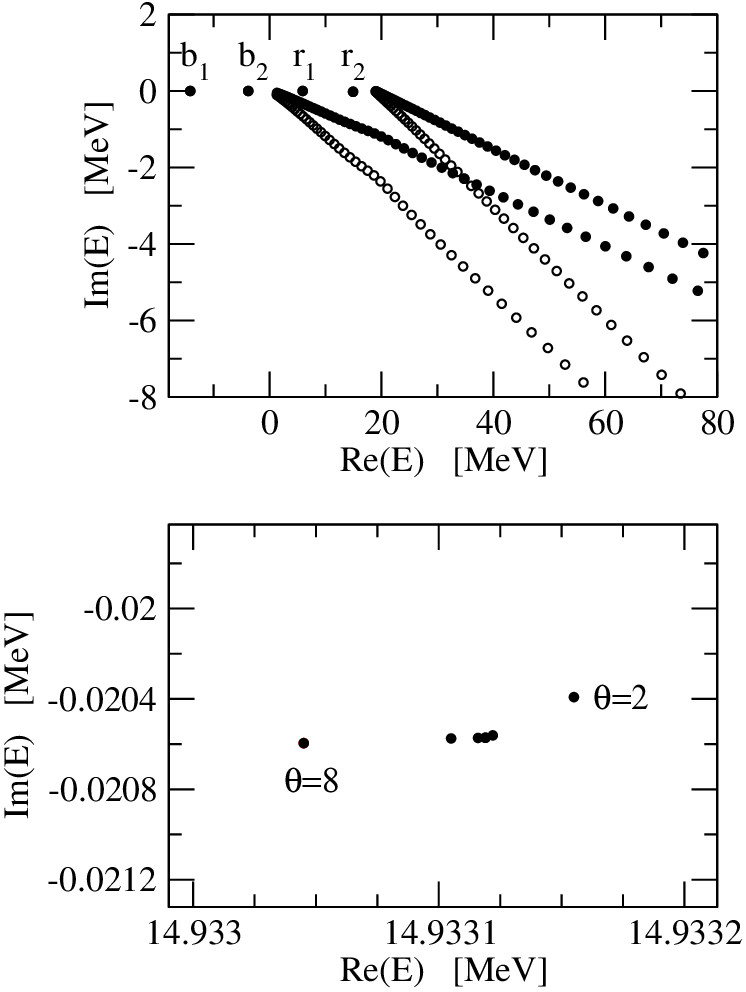}
\caption{Upper part: positions of the  $g_{9/2}$  states on complex $E$-plane 
with CS method with 
the rotation angle $\theta=2^o$  (filed circles) and with $\theta=4^o$ (
open circles) for the Coulomb potential Eq. (\ref{dilcoul}). The IAR is the state $r_2$. Lower part: the
vicinity of the IAR, the CS scaling parameter $\theta$ is varied between $2^o$ and
$8^o$ with a step size of $1^o$.}
\label{cspoles}
\end{figure}

From the mathematical theory of the complex scaling \cite{ABC1,ABC2,ABC3,ho83,moi98} it is known that
the continuous part of the spectrum of the complex scaled Hamilton operator
consist of half lines on the complex energy plane. The half lines  
start at the thresholds and they are rotated down from the real axis by $2\theta$.
In our calculation we have used $M_p=M_n=100$ basis functions and received
two hundred approximate complex 
eigenvalues from the diagonalization. These eigenvalues are plotted on Fig. 
\ref{cspoles} for two different 
$\theta$ values $\theta=2^o$ and $\theta=4^o$. From this Figure it is obvious
that the vast majority of the
eigenvalues correspond to discretization of the continuous spectrum. 
However there
are a few eigenvalues which are independent from the complex scaling parameter
$\theta$. These are denoted by letters $b_1$,$b_2$ and $r_1$,$r_2$ in the Figure \ref{cspoles}. 
The $r_2$ is the IAR
which is based mainly on the $2g_{9/2}$ bound neutron state as we have seen in the
CXSM calculation before. The $r_1$ resonance is based mainly on the narrow $2g_{9/2}$ proton resonance at
$E_i^{(p)}=(6.070,-2\times 10^{-5})$ MeV.
The bound states $b_1$ and $b_2$
originate on the $1g_{9/2}$ proton state at $E_i^{(p)}=-13.975$ MeV and the  $1g_{9/2}$ neutron state
at $E_i^{(n)}=-22.878$ MeV which are shifted up by
$\Delta_c=18.9$ MeV. 

\vskip 1.5cm

The lower part of
the Figure \ref{cspoles} show the so called $\theta$ trajectory i.e.  
the complex energy plane in the vicinity of the IAR when the complex scaling parameter changes between 
$\theta=2$ and $\theta=8$ degrees with step size of one degree. There is a small
change in the position and width of the resonance (this should be independent form the
value of $\theta)$ but this comes from the fact that a finite basis is used.
This phenomena is well know in all complex scaling calculation and there are
methods how to select the best approximation for the resonance \cite{yar78}.
The resonance position and width values given in Table \ref{comparecs} correspond to 
calculations with $\theta=4^o$.

A similarity of the CXSM and the CS method is that results become less accurate
if the contour of the CXSM or the rotated half lines are lying close
to the
resonance. To get high accuracy the resonance has to be well explored i.e.
should lie far above the contour. The rotated half lines of the CS play similar
role as the contours of the CXSM therefore we shall call the half lines
of the CS method also contours. Only the resonances above the contours can be
calculated.  This means that the $3g_{9/2}$ neutron resonance at
$E^{(n)}=(4.929,-6.035)$ $\arg (E^{(n)})=50.76^o$
 or the corresponding perturbed solution
can not be calculated by
the CS method since they can not be explored due to the critical angle
of the WS potential $\theta_{crit}=18.48^o$.
It can be calculated however by the CXSM using contours $LP3-LN3$ and we get for the perturbed energy
${\cal E}_p=(23.996,-6.147)$ MeV $\arg ({\cal E}_p-\Delta_c)=50.33^o$.

\begin{figure}[tbp]
\includegraphics[width=8.6cm]{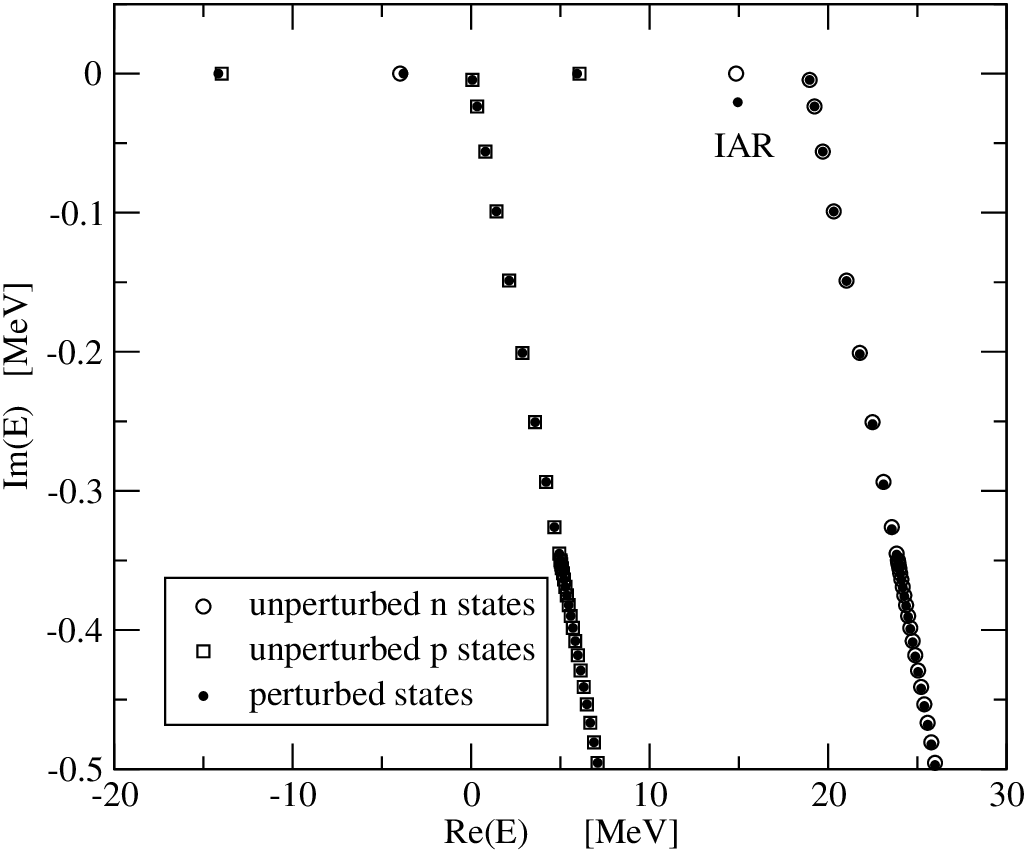}
\caption{Positions of the  $g_{9/2}$ Berggren basis states (circles for neutrons
and squares for protons) and the
results of the CXSM method (filled circles) on the
complex $E$-plane for the Coulomb potential Eq. (\ref{dilcoul}) with contours LP2 and LN2
in Table \ref{cont}. They are
similar to the optimal contour of the CS method.}
\label{mucspoles}
\end{figure}

The similarity of the methods can be seen even better if we try to use a contour
in CXSM, which resembles to the rotated continuum of the CS calculation.
In Fig. \ref{mucspoles} we present the results of the CXSM calculation
in which the contours LP2 and LN2 were chosen to be the same as the one corresponding to the optimal $\theta=4^o$
rotational angle of the CS calculation. One can see that the IAR is well
separated from the two contours starting at the origin and the one
starting at the neutron emission threshold. The  unperturbed pole
closest to the IAR is the bound $2g_{9/2}$ neutron state which is the dominating
component of the IAR wave function with amplitude: $C^{(n)}_i=(0.9918,-0.0043)$.
The second largest component of the IAR wave function is the one
of the $2g_{9/2}$ proton resonance with amplitude $C^{(p)}_i=(-0.1194,0.0001)$.
The wave function of the IAR practically unchanged as far as the discrete
components are concerned with respect to the case with the contours used
in Fig. \ref{poles} (LP1 and a real neutron contour). The energy of the IAR is ${\cal E}_p=(14.93309,-0.02058)$ MeV
coincide with the one ${\cal E}_p=(14.93303,-0.02062)$ MeV with the contours used
in Fig. \ref{poles} within the numerical error $1$ keV estimated from the
deviation from the CC results in Table \ref{comparecs}.
This good agreement convinces us that the use of the LP2 and LN2 contours
 which resembles to the contour of the CS
could also be used for calculating the IAR. 
The components of the different scattering states taken from the different
contours are certainly very different but the summed contribution
of the proton and neutron contours are basically the same. Since both are small
numbers their numerical values have little importance.
For the $g_{9/2}$ IAR the neutron continuum has negligible effect. For other partial waves
this effect is also small but not completely negligible.

An important difference between the results presented in Figs. \ref{cspoles} and
\ref{mucspoles} is that in Fig. \ref{cspoles} only perturbed states are shown
since in the CS method the basis states used are not eigenstates of any
unperturbed Hamiltonians. Therefore from the coefficients of the wave function
of the IAR explored we can not estimate the role of the unperturbed
neutron and proton states easily. To get similar quantities we have to calculate
the unperturbed state with the same CS contour and we have to calculate
overlaps with the IAR wave function.

\section{Summary}
Let us summarize briefly the results we received in this study.
We reproduced the results of the direct numerical solution of the coupled
Lane equations by diagonalizing the Hamiltonian in the full n-p  Berggren 
basis i.e. using the CXSM method.
The IAR parameters were extracted from the $S({\cal E}_p)$ calculated
by solving  the
Lane equation  along the real ${\cal E}_p$-axis by fitting it using the one pole 
approximation Eq. (\ref{poleform}). The fitted position $E_r(CC)$ and the width
$\Gamma(CC)$
of the IAR was compared to
the result of the CXSM calculation
and  the
agreement was generally better than 1 keV for all partial waves in which we
had IAR. In the wave function of the IAR furnished by the CXSM the 
contribution of the bound neutron
state has the dominant role and the proton resonance has a non negligible effect.
The integrated effect of the proton continuum is small but
essential to produce the correct width for the resonance. We studied the details of the different parts of the continuum segments and the
necessary numbers of the discretization points on the different segments.
The role of the cut off energy and the low energy part of the continuum
were also investigated. The neutron continuum played very small effect for the IAR-es.

The pole position of the IAR was calculated by complex scaling method as well.
For that we modified the Coulomb potential for a dilatational analytic one and
repeated the CC calculation and CXSM method with the modified
Coulomb potential. We received very good agreement to the numerical solution
of the coupled Lane-equations both with the CXSM  and the
CS methods. Therefore we conclude that in this case the CXSM
and the CS method give basically the same results apart some numerical errors which
naturally not the same in the two type of calculations.
This agreement suggests that the two method are basically equivalent in those cases
when both methods can be applied.

Besides the similarities and differences of the CXSM and the CS methods
discussed so far there are further important
differences between them. The application of the uniform CS method used here
is restricted to dilatation analytic potentials and the range of the rotational 
angle could also be limited. On the other hand in the CXSM method 
 the shape of the contour can be chosen with large flexibility although
 to go too deep into the complex energy might spoil a bit the accuracy of the
 calculated results. Another advantage of the CXSM is that the structure of the
 resonant state can be seen directly from the coefficients of the perturbed
 wave function. In the CS method the same information can be explored in a more
 indirect way.

 \acknowledgments Discussions with R. G. Lovas and R. J. Liotta are gratefully 
acknowledged. 
This work has been supported by PICT 21605 (ANPCyT-Argentina), by the 
Hungarian OTKA fund No. T46791, and by the Hungarian-Argentinian governmental 
fund NKTH Arg-6/2005-SECYT HU/PA05-EIII/005.

\end{document}